\documentclass[preprint,aps,pre,superscriptaddress,showpacs,floatfix]{revtex4}

\usepackage{graphicx}
\usepackage{amsmath}
\usepackage{epsfig}
\usepackage{natbib}

\begin{document}
\preprint{draft}
\title{Noise-to-Noise Ratios in Correlation Length Calculations Near Criticality}
\author{Avishay Efrat}
\affiliation{Physics Unit, Afeka Tel-Aviv Academic College of Engineering, Tel-Aviv 6910717, Israel}
\date{\today}
\begin{abstract}
For finite random systems, it is possible to define two types of variances (noises). It is demonstrated that their ratio is useful in calculating the correlation length of an infinite and rather general random system, as a function of temperature. The numerical method of obtaining those variables is not relevant. It can be real space numerical renormalization, simulation or any other method. It does not matter. The correlation length obtained by this novel technique may then be used to obtain directly the critical correlation exponent, $\nu$, rather than indirectly, using scaling relations, as is often done. The method is demonstrated by applying it to the random field Ising model.
\end{abstract}
\maketitle

Up until about a decade ago, quenched random systems were extensively studied, both theoretically and experimentally. Since then, the activity seems to have dwindled considerably. What may be interpreted as lack of interest due to that all the interesting problems have already been solved, reflects actually the difficulty of the field and lack of real progress. For example, even the values of the critical exponents at second order transitions are not known to good a accuracy after decades of research. Take, for instance, the Random Field Ising Model(RFIM) The various techniques used over the years for calculating the critical exponent, $\nu$, related to the correlation length, yielded results wide-ranging from 0.62 to 2.26. (The list of methods includes: Exact Ground State \cite{hy01,mf02,o86} Domain Wall Renormalization Group simulations \cite{c86}, Monte Carlo simulations \cite{r95,cm93,ry93,nb96,hd97}, Migdal Kadanoff Renormalization Group \cite{ry93,fbm95,fh9698}, Casher Schwartz Renormalization Group \cite{cs78,es03}, Modified Dimensional Reduction \cite{s88} and experiment \cite{bkj86,bcsy85,fhbh97,sbf99}).

In this paper, an effort is made to correct the situation by presenting a new method for calculating the correlation length, as a function of temperature, for rather general quenched random systems. The correlation length can also be obtained by measuring the correlation function \cite{es03}, but that requires much larger systems and many more calculations per each realization than the present technique. The reason for that difficulty is that it requires the calculation of the correlation function as a function of distance within each realization. It would also require many realizations to have enough statistics and to have a situation where the correlation length is still small compared to the size of the system, which becomes more problematic as the transition is approached \cite{es03}. Inspired by an argument first introduced by Brout \cite{b59}, the method is based on defining, for any local quantity of one's choice, two types of variances, which arise fundamentally from the two natural averaging procedures at our disposal: the average over the distribution of the randomness and the spatial average. It is shown how the ratio of two such variances yields directly the correlation length associated with the local quantity chosen. 'Local quantity' means here a function of a set of neighboring spins where the linear size of such a set is small compared to the correlation length. Note that for an infinite system, for any such linear size of the set, there is always a range of temperatures, close enough to the transition temperature, for which the correlation length is indeed much larger than the linear size of the set in mind. An immediate byproduct of calculating the correlation length as a function of temperature, close to the transition, is an estimate of the transition temperature, along with a direct measure of the critical exponent related to the correlation length.

The method presented here is quite general. It may be applied to any quenched random system. Any type of randomness may be considered and variances of any local physical quantity may be used. Also, the way the variances are obtained is actually irrelevant and any numerical method for obtaining those variances will do. Due to its generality, the method may be expected to provide a novel useful tool in the study of quenched random systems. As will be seen in the following, the discussion assumes Ising systems. A careful examination of the derivation of the main results will convince the reader that the results obtained are more general. To give a concrete demonstration, the method is applied to the spin-spin correlation function of the three-dimensional RFIM.

Consider a large, but finite, general quenched random system of Ising spins, represented by the random Hamiltonian
\begin{equation} 
{\cal H}=-{\cal H}_0-\sum_{A}h_AS_A.
\label{EqHam}
\end{equation}
where the coupling constants in the ferromagnetic Hamiltonian, ${\cal H}_0$, are position independent and the sum over $A$ is over local subsets of neighboring spins in the system. The notation $S_A$ is used for the product of all spins belonging to a subset $A$, 
\begin{equation} 
S_A = \prod_{i\in A} s_i.
\label{EqSA}
\end{equation}
The $h_A$'s are random couplings, which average to zero but may be short ranged correlated. Here are two examples to clarify the above. For ${\cal H}_0$ the nearest-neighbors ($nn$) ferromagnetic Ising Hamiltonian, the traditional random field Ising model is obtained by choosing the sets $A$ to be single sites. The random bond Ising model is obtained by choosing those sets to be pairs of $nn$.

Let us denote by $X_{\boldsymbol k}$ a local quantity associated with a location $\boldsymbol k$ on a lattice of volume (the total number of sites), $V$. The number of locations $K$ is of the order of $V$, but it is usually larger. For example, if $X_{\boldsymbol k}$ is taken to be the spin at $\boldsymbol k$, the total number of locations, $K$, is $V$. If $X_{\boldsymbol k}$ is chosen as the product of a pair of $nn$ spins, the locations $\boldsymbol k$ correspond to the centers of the bonds connecting the two spins in each pair and thus $K=Vd$ on a d-dimensional hyper cubic lattice. Define next $x_{n\boldsymbol k}$ to be the thermal average of $X_{\boldsymbol k}$, taken for a given realization of the randomness, $n$, with $n=1,\ldots,N$. 

Define next two types of averages along with their relevant variances. The first is the spatial average of $x_{n\boldsymbol k}$ within some realization $n$ of the randomness, that is,
\begin{equation}  
\bar x_n \equiv \frac{1}{K} \sum_{\boldsymbol k}x_{n\boldsymbol k},
\label{EqSpatialAv}
\end{equation}
Its squared variance is given by
\begin{equation} 
\varepsilon^2_{xn} \equiv \frac{1}{K} \sum_{\boldsymbol k}x^2_{n\boldsymbol k}-\bar x^2_n.
\label{EqEpsilonRealSqrd}
\end{equation}
The second is the total average, the average of $\bar x_n$ over all realizations:
\begin{equation}  
\bar x \equiv \frac{1}{N} \sum_{n=1}^N \bar x_n=\frac{1}{NK} \sum_{n=1}^N \sum_{\boldsymbol k} x_{n\boldsymbol k},
\label{EqTotalAv}
\end{equation}
To the above total average, one can associate two types of variances. The first is related to the single summation representation of $\bar x$ in Eq. (\ref{EqTotalAv}) and is defined as
\begin{equation} 
\delta^2_x \equiv \frac{1}{N} \sum_{n=1}^N \bar x^2_n-\bar x^2.
\label{EqDeltaSqrd}
\end{equation}
The second is obviously related to the double summation representation of $\bar x$ in Eq. (\ref{EqTotalAv}) and is defined as
\begin{equation} 
\sigma^2_x \equiv \frac{1}{NK} \sum_{n=1}^N \sum_{\boldsymbol k} x^2_{n\boldsymbol k}-\bar x^2.
\label{EqSigmaSqrd}
\end{equation}
Note that these two variances are not the same since
\begin{equation} 
\sigma^2_x-\delta^2_x = \frac{1}{N} \sum_{n=1}^N \left(\frac{1}{K} \sum_{\boldsymbol k} x^2_{n\boldsymbol k}-\bar x^2_n\right) = \frac{1}{N} \sum_{n=1}^N \varepsilon^2_{xn} \equiv \varepsilon^2_x.
\label{EqEpsilonSqrd}
\end{equation}

\begin{figure}[ht]
\includegraphics[width=.48\textwidth]{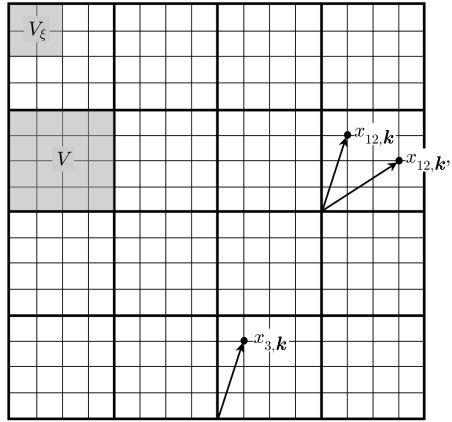}
\caption{\label{Fig1} 
\footnotesize{A two-dimensional representation is given schematically. Each square confined by thick lines is a different random realization of the system. The volume $V$ is the number of sites in a square of linear size $L$. The correlation volume $V_\xi$ is the correlation length squared. Thick arrows represent location vectors $\boldsymbol k$. The number of locations, $K$, is taken to be equal to the system's size, $V$.}}
\end{figure}

Let $K_\xi$ be the number of locations, $\boldsymbol k$, within a correlation volume, at which $x_{n\boldsymbol k}$ is obtained. The geometry described above is represented in Fig. \ref{Fig1}, for a two-dimensional system. Systems of linear size $L$ corresponding to different realizations are depicted side by side.

\textbf{\textit{Away from the transition}}, where the correlation length, $\xi$, is much smaller than the linear size of the system, $L$, the number of statistically independent variables in the system, $K/K_\xi$ , is very large (note that for the RFIM this is possible only for temperatures above the transition, because below the transition the correlation length is always of the order of $L$ \cite{es03,dsy93}). In that case, according to the central limit theorem, $\bar x_n$ may be viewed as distributed normally around its average, $\bar x$, with variance $\sigma_x^2/(K/K_\xi)$. Following the definition of $\delta_x^2$, given by Eq. (\ref{EqDeltaSqrd}), one obtains $\delta^2_x = \sigma^2_x\frac{K_\xi}{K}$, or
\begin{equation} 
K_\xi = \left(\frac{\delta_x}{\sigma_x}\right)^2 K,
\label{EqDeltaOverSigma}
\end{equation}
For the following, it proves useful to define, for the general local quantity $x$, a function
\begin{equation} 
f_x(T,L) \equiv \left(\frac{\delta_x(T)}{\sigma_x(T)}\right)^{2/d} L,
\label{EqF}
\end{equation}
where $T$ is the temperature and $d$ is the dimensionality of the system. For example, if $x$ represents the local magnetization, $m$, or the local susceptibility, $\chi$, it is clear that $K=L^d$ and $K_\xi=\xi^d$, since both quantities are defined for a single site. Thus, when $\xi$ is much smaller than $L$, substituting the above $K$ and $K_\xi$ into Eq. (\ref{EqDeltaOverSigma}), leads to
\begin{equation} 
\xi(T,L)=f_x(T,L).
\label{EqFeqXi}
\end{equation}
It may seem that Eq. (\ref{EqFeqXi}) is based on the local quantities being just single site quantities. The interested reader could, however, convince himself easily that it holds for any local quantity, as long as the linear extent of that quantity is considerably smaller than the correlation length.

\textbf{\textit{Close to the transition}}, where the correlation length, $\xi$, is close to the linear size of the system, $L$, the number of independent variables in the system, $K/K_\xi$ , tends to unity. Therefore, the central limit theorem does not apply anymore and so does not Eq. (\ref{EqDeltaOverSigma}). In that case, one cannot assume, using Eq. (\ref{EqDeltaOverSigma}), that necessarily $\delta_x/\sigma_x$  tends to unity nor use Eq. (\ref{EqFeqXi}) to obtain $\xi$ .The question arising now is how to proceed from here to obtain the correlation length critical exponent, $\nu$. The possible difficulties and a practical strategy for performing that task are outlined in the following paragraphs.

\textbf{\textit{Regarding the infinite system}}, an obvious but important observation is that for any temperature above $T_c$, no matter how close it is to $T_c$, the correlation length is finite and therefore infinitely small compared to the size of the system. It means that, regarding the infinite system, Eq. (\ref{EqFeqXi}) holds for any temperature $T$, above $T_c$. It is also clear that for finite systems, there should always be a region of temperatures, above $T_c$, where $\xi$ is independent of $L$. Therefore, for that region, one can use $f_x(T,L)$, of that finite system, to describe $\xi$ of the infinite system. It also means that the validity of that description extends to temperatures closer and closer to $T_c$ as the system size is increased accordingly. In Eq. (\ref{EqFeqXi}) the temperature dependence of $\xi$ enters via the temperature dependence of $\delta$ and $\sigma$. It may be expected, therefore, that for a set of finite sizes, $\{L_i\}$ all the functions of temperature, $f_x (T,L_i)$, will merge with $\xi(T,\infty)$ for temperatures away from $T_c$. For temperatures closer to $T_c$, however, these functions are expected to depart from one another and from the infinite system correlation length according to system size.

The asymptotic behavior of $\xi(T)$ near the transition temperature, $T_c$, defines the critical exponent $\nu$, by
\begin{equation} 
\xi(T)\approx A|T-T_c|^{-\nu},
\label{EqXiAsymptotic}
\end{equation}
where $A$ is a constant. Very close to $T_c$, it is clearly impossible for $\xi$ to be described by Eq. (\ref{EqFeqXi}). It becomes possible somewhat away from $T_c$, provided the size of the system is large enough. What is needed is a range of temperatures for which Eqs. (\ref{EqFeqXi}) and (\ref{EqXiAsymptotic}) will both hold. In that range, which is possible to find in principle, by increasing the size of the system, one can equate $\xi(T)$ of Eq. (\ref{EqFeqXi}) with that of Eq. (\ref{EqXiAsymptotic}) to obtain
\begin{equation} 
\ln\left[\left(\frac{\delta_x(T)}{\sigma_x(T)}\right)^{2/d} L\right]\approx -\nu \ln |T-T_c|+\ln A,
\label{EqLnXiAsymptotic}
\end{equation}
It may seem unclear, though, whether systems large enough are actually practical. There is a reason to believe, however, that indeed, such temperature regions where both Eqs. (\ref{EqFeqXi}) and (\ref{EqXiAsymptotic}) hold simultaneously exist. Prior experience with RFIM \cite{es03} suggests that large enough systems are numerically accessible. Although the technique of obtaining the correlation length there is less efficient than the technique presented here, the task of obtaining the critical exponent $\nu$ faces the same difficulties.

\textbf{\textit{As a demonstration of the method}}, consider the random field Ising Hamiltonian,
\begin{equation} 
{\cal H}=-J\sum_{<i,j>}s_is_j-\sum_i h_is_i,
\label{EqRFHam}
\end{equation}
where the pair $<ij>$ denotes a $nn$ pair on a cubic lattice with periodic boundary conditions. The $h_i$'s are random uncorrelated fields, distributed around zero,
\begin{equation} 
\overline{h_i}=0, \qquad \overline{h_i h_j} = h^2\delta_{ij},
\label{EqMeanStndrdDev}
\end{equation}
where $\delta_{ij}$ is the Kronecker delta. The bar denotes ensemble average and the random fields are distributed according to a Gaussian distribution,
\begin{equation} 
P\{h\} = \prod_i P_i \{h_i\} =
\frac{1}{(h\sqrt{2\pi})^K}exp\left(-\frac{1}{2h^2}\sum_i{h_i}^2\right),
\label{EqGaussDist}
\end{equation}
Take the local quantity,
\begin{equation} 
X_i=\frac{s_i}{h^2}\sum_{j=1}^K h_j.
\label{EqXi}
\end{equation}
I chose to multiply the local dynamical variable $s_i$ by  $\sum_{j=1}^K h_j/h^2$, because its thermal and ensemble average has a very interesting meaning, namely,
\begin{equation} 
\overline{\langle X_i\rangle} = \beta\sum_{j=1}^K \overline{\langle s_is_j \rangle -
\langle s_i\rangle\langle s_j\rangle}.
\label{EqSusc}
\end{equation}
Since $\overline{\langle X_i \rangle}$ does not depend on $i$, it is easy to show \cite{es03,dsy93,ss8586} that it is exactly the ensemble averaged susceptibility. (Note that the form of the distribution (\ref{EqGaussDist}) is essential in deriving the exact relation (\ref{EqSusc}).)

Both, $\delta_x(T)$ and $\sigma_x(T)$, were calculated for $x_i=\langle X_i \rangle$, using their definitions given by Eqs. (\ref{EqDeltaSqrd}) and (\ref{EqSigmaSqrd}). The procedure I use here for obtaining $x_i$ numerically, is based on the Casher-Schwartz RSRG \cite{cs78,es03,dsy93,sf80}. As other real space techniques, the Casher-Schwartz RSRG provides simple, one step, recursion relations for the fields and couplings constants. A set of $V=L^3$ Ising spins, with $L=2^n$, situated on a three-dimensional cubic lattice with periodic boundary conditions, is considered. First a realization of the distribution (\ref{EqGaussDist}) is generated. Then the Casher-Schwartz procedure is applied $n-1$ times. This results in a cubic system of the $2\times2\times2$ spins surviving the procedure. The smallness of the system allows brute force calculation of the remaining $8$ $\langle s_i \rangle$'s. The ''sites translation'' method \cite{es03} is used to enable the calculation of the thermal average of all the spins in the original lattice. The actual ensemble averages were performed by repeating the calculations for $10,000$ realizations of the randomness. The averages obtained for each realization are distributed with the variance $\delta_x$ as defined by Eq. (\ref{EqDeltaSqrd}), while the local quantity itself is distributed over the entire system (including all realizations) with the variance $\sigma_x$ as defined by Eq. (\ref{EqSigmaSqrd}). Finally, the numerical procedure described above is repeated for different temperatures, while setting $h=T/2$. The largest system considered here is $L=64=2^6$.

\begin{figure}[ht]
\includegraphics[width=.48\textwidth]{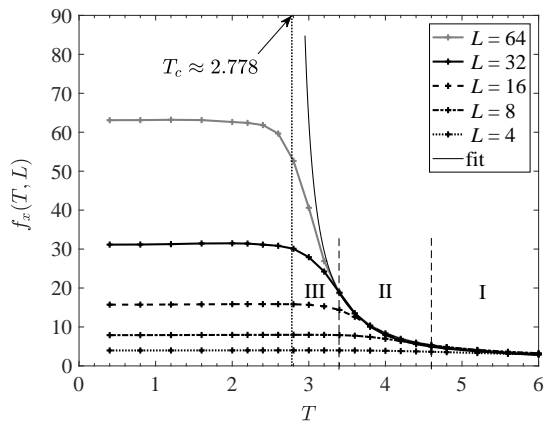}
\caption{\label{Fig2} 
\footnotesize{The function $f_x(T,L)$, is plotted vs. temperature. Also shown is a fit of the asymptotic behavior of $\xi(T)$ to $f_x(T,L)$ of the largest $L=64$ system. It is clearly seen how the lines of the finite systems depart from $\xi(T)$ depending on system size.}}
\end{figure}

In Fig. \ref{Fig2} the function $f_x(T,L)$ is plotted vs. temperature. All quantities presented in Fig. \ref{Fig2} are dimensionless. Length is measured in units of lattice constant. Taking the spin variables to be dimensionless, makes the parameters $h$ and $J$ in the Hamiltonian (\ref{EqRFHam}) have the dimensions of energy. To use in the figure dimensionless quantities, $h$ and $T$ are rescaled: $h/J \rightarrow h$ and $kT/J \rightarrow T$. For $L = 64$, three temperature ranges can be observed in the figure: \textbf{I.} Far from $T_c$, where $(\delta/\sigma)^{2/3}L=\xi$ is not expected to behave as $(T-T_c)^{-\nu}$. \textbf{II.} The temperature range used for the fit, where $(\delta/\sigma)^{2/3}L=\xi\propto(T-T_c)^{-\nu}$. \textbf{III.} Too close to $T_c$, where $\xi\propto(T-T_c)^{-\nu}$, yet $(\delta/\sigma)^{2/3}L\ne\xi$. To establish the critical temperature, $T_c$, of the infinite system, along with the critical exponent, $\nu$, the asymptotic behavior of $\xi(T)$ as given by Eq. (\ref{EqXiAsymptotic}) was fitted to the $(\delta/\sigma)^{2/3}L$ line of the largest $L=64$ system, using only the range of temperatures, denoted as II in the figure. The proper temperature range II was chosen as the one giving the best goodness of the fit. The resulting fit is also shown in Fig. \ref{Fig2}. The resulting fit parameters are:
$T_c=2.778 (2.715,2.84)=2.778\pm0.063$, 
$\nu=1.178 (1.098,1.258)=1.178\pm0.080$, 
$A=10.67 (9.836,11.5)=10.67\pm0.834$, 
with goodness parameters:
Adjusted R-square: $0.9999$, 
RMSE: $0.03876$.

\begin{figure}[ht]
\includegraphics[width=0.48\textwidth]{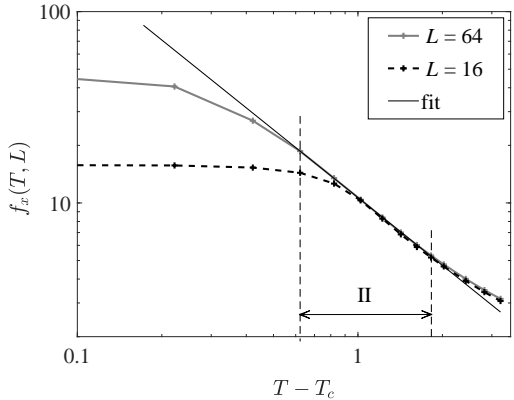}  
\caption{\label{Fig3} 
\footnotesize{A Log-Log plot of Fig. \ref{Fig2}. The temperature range, II, used for the fit is the one where $ln[(\delta_\xi/\sigma_\xi)^{2/3}L]$ is indeed linear with $ln|T-T_c|$. For the sake of clarity only two system sizes are presented. The slope is obviously $-\nu$.}}
\end{figure}

In Fig. \ref{Fig3}, a Log-Log plot of Fig. \ref{Fig2} is presented. This is to demonstrate that the temperature range used for the fit is the one where $ln[(\delta_\xi/\sigma_\xi)^{2/3}L]$ is indeed linear with $ln|T-T_c|$.

The resulting $\nu=1.178\pm0.080$ obviously agrees with part of the previous results cited at the onset of this paper (Exact Ground State \cite{hy01,mf02}, Domain Wall Renormalization Group simulations \cite{c86}, Monte Carlo simulations \cite{r95} and experiment \cite{bkj86}). Being a totally independent and direct determination, the present result gives credence to those results over others.

To summarize, in this article I have presented a new and most general approach for calculating the correlation length of an infinite general random system, as a function of temperature, using noise-to-noise ratios. It is also outlined how the attained dependence of the correlation length on temperature can be cautiously used to obtain the correlation length critical exponent $\nu$. The applicability of the method has been demonstrated on the random field Ising model. This should not obscure the generality of the method, as it may be applied to any quenched random system, under various types of randomness. Also, variances of any local physical quantity, obtained by any numerical method, may be used.  I expect to apply it to other systems in the near future. Hopefully, the present article, will enable others, who may use other techniques for numerical study of quenched random systems, to obtain the critical correlation length exponent by using noise-to-noise ratios.



\end{document}